\documentclass{PoS}
\newcommand{\Bsmumu}{$B_s\to\mu^+\mu^-$}
\newcommand{\Bdmumu}{$B_d\to\mu^+\mu^-$}
\newcommand{\MeVcc}{\ensuremath{\,{\rm MeV}/c^2}\xspace}
\newcommand{\BRof}[1]{{\ensuremath{\cal B}}(#1)}
\newcommand{\CL}{CL\ }

\newcommand{\CLs}{\ensuremath{\textrm{CL}_{\textrm{s}}}\xspace}
\newcommand{\CLb}{\ensuremath{\textrm{CL}_{\textrm{b}}}\xspace}

\newcommand{\BR} {{\ensuremath{\cal B}\xspace}}

\newcommand{\invfb} {\ensuremath{\mbox{\,fb}^{-1}}\xspace}
\newcommand{\BuToJPsiK} {\decay{\Bu}{\jpsi K^+}\xspace}

\newcommand{\BsKK}{\ensuremath{\Bs\to K^+K^-}\xspace}

\newcommand{\BdKpi}{\ensuremath{\Bd\to K^+\pi^-}\xspace}

\newcommand{\Bhh}{\ensuremath{B^0_{(s)}\to h^+h^-}\xspace}
\newcommand{\Bmm}{\ensuremath{B^0_{(s)}\to \mu^+\mu^-}\xspace}

\newcommand{\BsToJPsiPhi}  {\decay{\Bs}{\jpsi\phi}\xspace}
\newcommand{\BdToKpi}      {\decay{\Bd}{K^+\pi^-}\xspace}
\newcommand{\jpsi}{\ensuremath{J/\psi}\xspace}
\newcommand{\Bd}      {\ensuremath{B^0}\xspace}
\newcommand{\Bs}      {\ensuremath{B^0_s}\xspace}
\newcommand{\Bu}      {\ensuremath{B^+}\xspace}

\newcommand{\decay}[2]{\ensuremath{#1\!\to #2}\xspace}    
\usepackage{xspace}  
\usepackage{rotating}
\usepackage{multirow}
\usepackage{placeins}
\usepackage{amssymb}
\usepackage{amsfonts}
\usepackage{upgreek}
\usepackage{epsfig}
\title{Search for the rare decays $B^0_{(s)}\to \mu^+\mu^-$  at LHCb}

\ShortTitle{Search for $B^0_{(s)}\to \mu^+\mu^-$  at LHCb}

\author{\speaker{Justine Serrano}\thanks{On behalf of the LHCb collaboration}\\
        CPPM, Aix-Marseille Universit\'e, CNRS/IN2P3, Marseille, France\\
        E-mail: \email{serrano@cppm.in2p3.fr}}

\abstract{A search for the  \Bsmumu\ and \Bdmumu\ decays is presented
using $\sim 300$\,pb$^{-1}$ of $pp$ collisions at $\sqrt{s}$ = 7\,TeV
collected by the LHCb experiment at the Large Hadron Collider at CERN.  
The measured upper limit for the branching ratio of the  \Bsmumu\ decay is \BRof \Bsmumu$< 1.3  \ (1.6) \times 10^{-8}$ at 90\,\% (95\,\%) confidence level (CL), 
while in the case of the \Bdmumu\ decay the measured upper 
limit is \BRof \Bdmumu $< 4.2 \ (5.1) \times 10^{-9}$ at 90\,\% (95\,\%) CL.
A combination with the  2010 dataset results 
in \BRof \Bsmumu $<1.2 \ (1.5) \times 10^{-8}$ at 90\,\% (95\,\%) CL.
}         

\FullConference{The 2011 Europhysics Conference on High Energy Physics-HEP 2011,\\
		July 21-27, 2011\\
		Grenoble, Rh\^one-Alpes France}

\begin{document}
\section{Introduction}
Measurements at low energies may provide interesting indirect constraints
on the masses of particles that are too heavy to be produced directly.
This is particularly true for Flavour Changing Neutral Currents
(FCNC) processes which are highly suppressed in the Standard Model (SM) 
and can only occur through higher order diagrams. 
The SM prediction for the branching ratios (\BR) of the FCNC decays \Bsmumu\
and \Bdmumu\ have been computed  to be  $(3.2 \pm
0.2) \times 10^{-9}$ and  $(0.10 \pm 0.01) \times 10^{-9}$ respectively~\cite{Buras2010}.
However New Physics (NP) contributions can significantly enhance these values.

The best published limits from the Tevatron at $95\%$~\CL  
are obtained using 6.1 fb$^{-1}$ by the D0 collaboration~\cite{d0_PLB}, 
and using 2 fb$^{-1}$ by the CDF collaboration~\cite{cdf_PRL}. 
The CDF collaboration has also presented a
preliminary result~\cite{cdf_bsmumu_preprint} with 6.9 fb$^{-1}$
in which an excess of \Bsmumu candidates 
is reported, compatible with a \BRof \Bsmumu = $(1.8^{+1.1}_{-0.9}) \times 10^{-8}$.

The LHCb collaboration has previously obtained the limits 
\BRof \Bsmumu $<5.4 \times 10^{-8}$ and  \BRof \Bdmumu$<1.5 \times 10^{-8}$ at 95$\%$ \CL
based on 37 pb$^{-1}$ of luminosity collected in the 2010 run \cite{LHCb_paper}. 
We present here a measurement based on 
300 pb$^{-1}$ of integrated luminosity collected between March and June 2011.

\section{Analysis strategy}

The general structure of the analysis is similar to the one described 
in Ref.~\cite{LHCb_paper} and is detailed in Ref.~\cite{LHCb_conf}.

The selection procedure 
treats signal and control/normalization channels in the same way
in order to minimize the systematic uncertainties.
Assuming the SM branching ratio and the $b \bar{b}$ cross-section, measured
within the LHCb acceptance, of $\sigma_{b \overline{b}} = 75\pm14\,\mu$b~\cite{bbxsection}, 
approximately $3.4~(0.32)$ 
$B^0_{s}\to \mu^+\mu^-$ ($B^0 \to \mu^+\mu^-$)  events
are expected to be reconstructed and selected in the analysed sample.

After the selection, each event is given a probability to be signal or background in a 
two-dimensional space defined by two independent likelihoods: 
the invariant mass and the output of a Boosted Decision Tree (BDT) from the TMVA package~\cite{tmva}.
The combination of variables entering the BDT is optimized using Monte Carlo (MC)
simulation. The following variables have been used: the $B$ lifetime, impact parameter and transverse momentum of the $B$, the  minimum impact parameter significance of the muons, the distance of closest approach between the two muons, the degree of isolation of the two muons with respect to any other track in the event, the cosine of the polarization angle, the $B$ isolation and the minimum $p_T$ of the muons. The BDT distribution is then transformed in order to be flat for the signal and peaked at 0 for the background.

\begin{figure}[t]
\centering
\includegraphics[width=.55\textwidth]{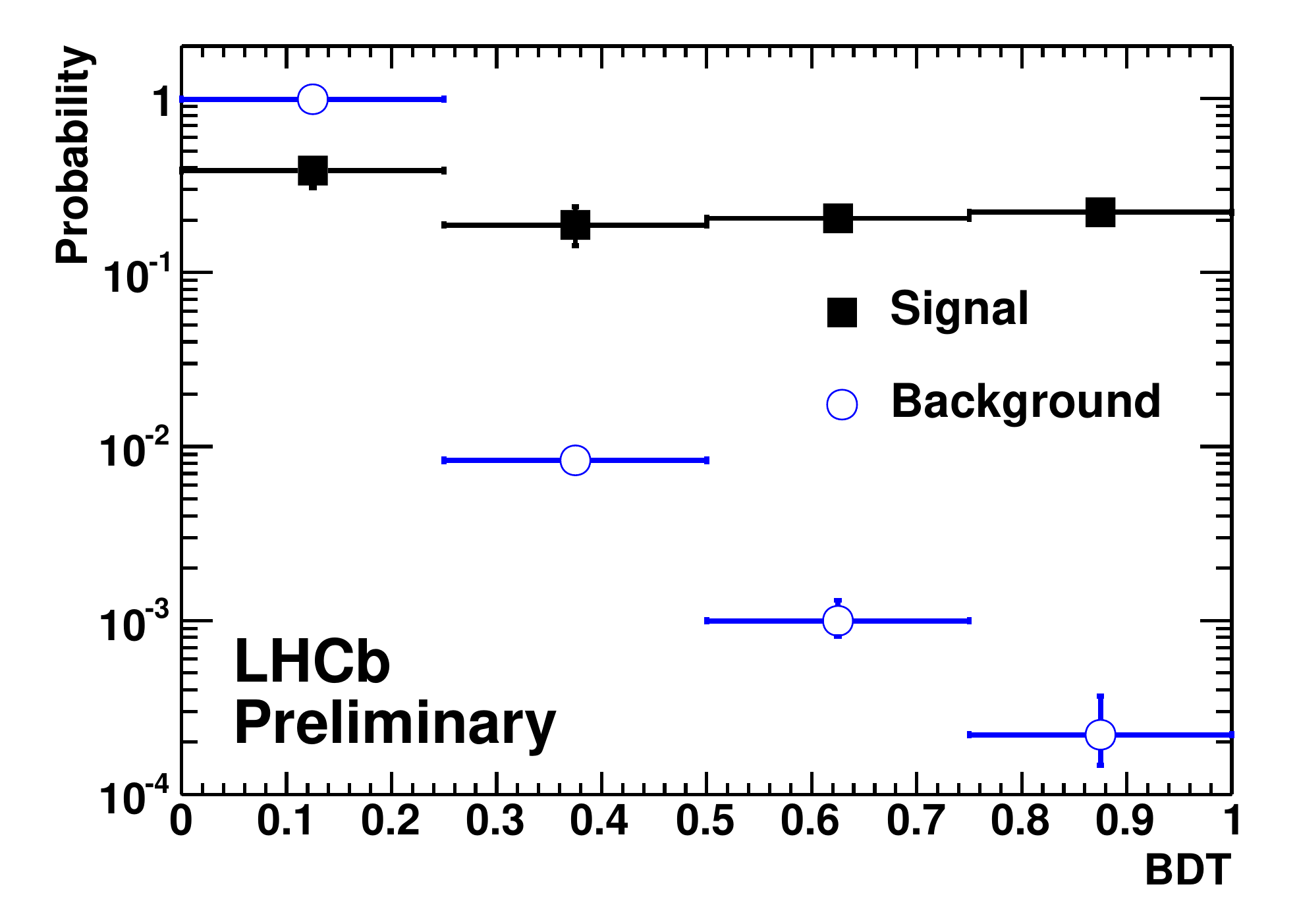}
\vspace{-3mm}
\caption{BDT calibration for signal and background.}
\label{fig:BDT_all}
\end{figure}

The calibration of the invariant mass and the BDT likelihoods  are obtained from data using  control samples.
The signal BDT shape is obtained from \Bhh events free from trigger biases while 
the background shape is obtained using sideband \Bmm candidates.
The resulting distributions are shown in Fig.~\ref{fig:BDT_all}.

The parameters describing the
invariant mass line shape  of the signal are extracted from data using control samples.
The average mass values are obtained from \BdKpi and \BsKK exclusive samples. The \Bs and \Bd
mass resolutions
are estimated by interpolating the ones 
obtained with the dimuon 
resonances ($J/\psi, \psi(2S)$ and $\Upsilon(1S,2S,3S)$) and cross-checked via a fit
to the invariant mass distribution of the \Bhh inclusive decays and of the \BdKpi exclusive decay. The interpolation yields 
 $\sigma(B) = 24.6 \pm 0.2_{\rm stat} \pm 1.0_{\rm syst}$.

The number of expected signal events is obtained by normalizing 
to channels of known branching ratios, 
\BuToJPsiK, \BsToJPsiPhi, and \BdToKpi, that
are selected in a way as similar as possible to the signal.

The probability for a background event to have a given BDT and invariant 
mass value is obtained by a fit of the mass distribution of events in the mass sidebands, in bins of BDT. Different fit functions and mass ranges are used to compute the systematics uncertainties.
The two-dimensional space formed by the invariant mass  
and BDT is binned, and for each bin we compute how many events 
are observed in data, how many signal events are expected for a 
given \BR \xspace hypothesis and luminosity, 
and how many background events are expected for a given luminosity.
The compatibility of the observed distribution of events in
all bins with the expected one for a given \BR\xspace hypothesis 
is computed using  
the CLs method~\cite{Read_02}, which allows to exclude a given hypothesis at a given confidence level.

In order to avoid unconscious biases, 
the data in the mass region defined by $M_{\Bd} - 60 \MeVcc $ 
and  $ M_{\Bs} + 60 \MeVcc$ have been blinded until the completion of the analysis.

\section{Results}

\begin{figure}[tbp]
  \begin{center}
     \includegraphics*[width=0.55\textwidth]{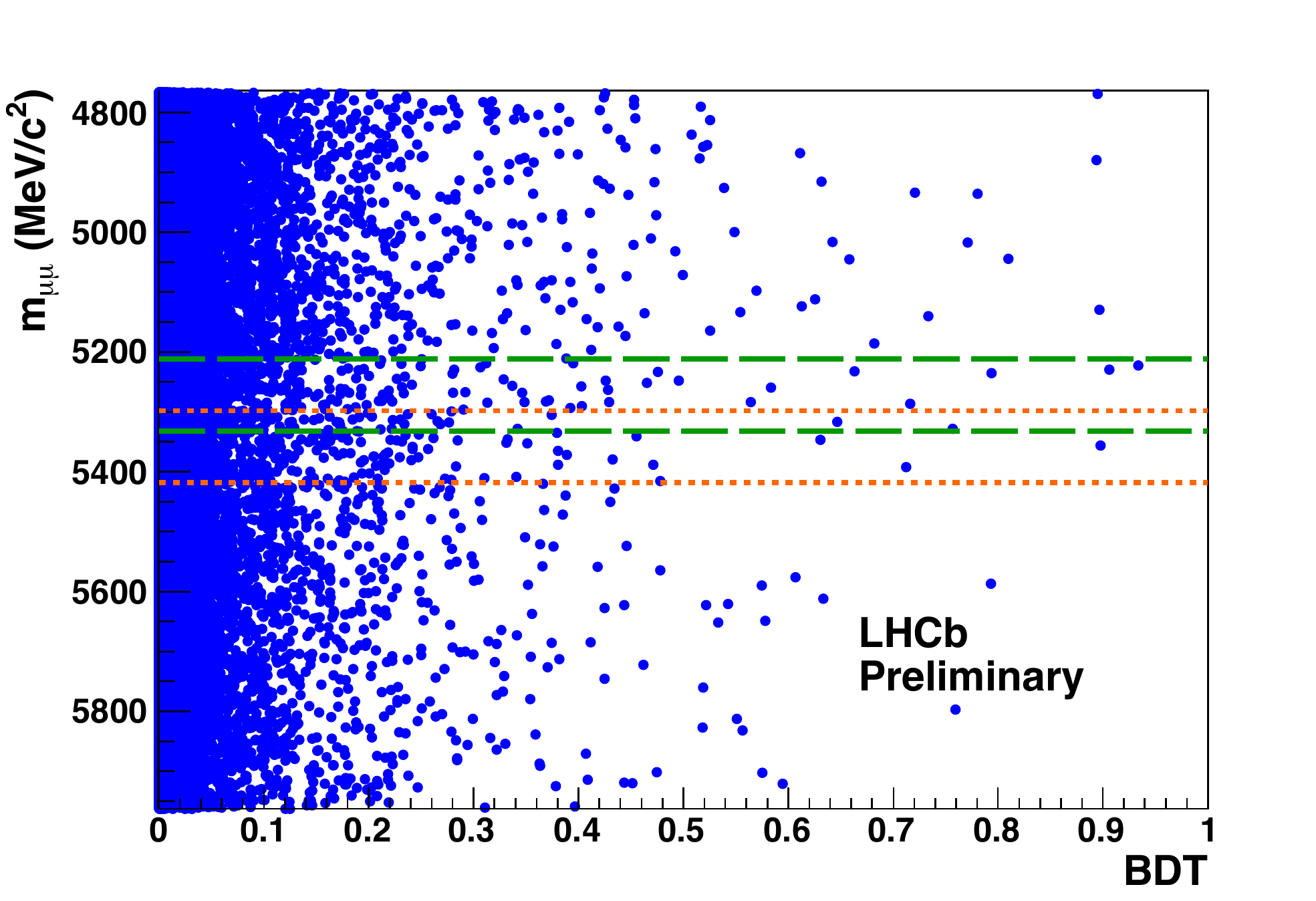}
  \end{center}
\vspace{-3mm}
  \caption{Distribution of selected dimuon events in the 
     invariant mass {\it vs} BDT plane. The orange short-dashed 
(green long-dashed) lines indicate the $\pm 60 \MeVcc$ search window around
    the \Bs (\Bd).}
\label{fig:BDTvsmass_obs}
\end{figure}

The distribution of events in the invariant mass versus BDT 
plane is reported Fig.~\ref{fig:BDTvsmass_obs}.
The expected limit at 90 (95)\,\% CL for the \Bsmumu\ is $ 0.8~(1.0) \times 10^{-8}$ in the case of background only hypothesis. When adding signal events according to the SM branching fraction, these limits become 
 $ 1.2~(1.5) \times 10^{-8}$.
The observed values for the \Bsmumu channel is $ 1.3~(1.6) \times 10^{-8}$ with a CLb value of 0.80.
The observed events  are in good agreement with the background expectations and the presence of \Bsmumu events according to SM predictions.

For the \Bdmumu, the expected limit at 90 (95)\,\% CL  is $ 2.4~(3.1) \times 10^{-9}$ in the case of background only hypothesis. 
The observed values is $ 4.2~(5.2) \times 10^{-9}$ with a CLb value of 0.79.
The comparison of the observed distribution of events with the expected background distribution 
results in a  p-value (1-\CLb) of 20\,\% (21\,\%) for the \Bsmumu\ (\Bdmumu) decays.
In the case of \Bdmumu, the slightly low p-value is due to an excess of the observed 
events in the most sensitive BDT bin with respect to the background expectations. 
A larger data sample will 
allow to clarify the situation. 
In the case of  \Bsmumu, when a signal is included at the level 
expected in the Standard Model, the p-value increases to 50\,\%.

Finally, the \Bsmumu\ limit is combined with the one published from the 2010 data to obtain
\BRof{\Bsmumu} < $1.2~(1.5) \times 10^{-8}$ at 90\,\% (95\,\%) CL. This 90\,\% CL upper limit is still
3.8 times above the standard model prediction.

\end{document}